 \input pictex.sty   
\input dcpic.sty
%
%

%
%
%
%

\def\Serif{cmr}
\def\SerifBold{cmbx}
\def\SerifItalics{cmti}
\def\SerifSlanted{cmsl}
\def\SerifBoldItalics{cmbxti}
\def\SansSerif{cmss}
\def\SansSerifBold{cmssbx}
\def\SansSerifItalics{cmssi}
\def\SansSerifSlanted{cmssi}
\def\Math{cmmi}
\def\Symbols{cmsy}
\def\MathBold{cmmib}
\def\MoreSymbols{cmex}
\def\Typewriter{cmtt}
\def\Gothic{eufm}
\def\Double{msbm}

= 			\Serif10 			at 5pt
= 		\SerifBold10 		at 5pt
= 	\SerifItalics10 	at 5pt
=		\SerifSlanted10 	at 5pt
=	\SerifBoldItalics10	at 5pt
= 		\SansSerif10 		at 5pt
=	\SansSerifBold10	at 5pt
=	\SansSerifItalics10	at 5pt
=	\SansSerifSlanted10	at 5pt
=				\Math10				at 5pt
=			\MathBold10			at 5pt
=			\Symbols10			at 5pt
=		\MoreSymbols10		at 5pt
=		\Typewriter10		at 5pt
=			\Gothic10			at 5pt
=			\Double10			at 5pt

= 			\Serif10 			at 7pt
= 		\SerifBold10 		at 7pt
= 	\SerifItalics10 	at 7pt
=	\SerifSlanted10 	at 7pt
=\SerifBoldItalics10	at 7pt
= 		\SansSerif10 		at 7pt
= 	\SansSerifBold10 	at 7pt
=\SansSerifItalics10	at 7pt
=\SansSerifSlanted10	at 7pt
=			\Math10				at 7pt
=		\MathBold10			at 7pt
=			\Symbols10			at 7pt
=		\MoreSymbols10		at 7pt
=		\Typewriter10		at 7pt
=			\Gothic10			at 7pt
=			\Double10			at 7pt

= 			\Serif10 			at 8pt
= 		\SerifBold10 		at 8pt
= 	\SerifItalics10 	at 8pt
=	\SerifSlanted10 	at 8pt
=\SerifBoldItalics10	at 8pt
= 		\SansSerif10 		at 8pt
= 	\SansSerifBold10 	at 8pt
=\SansSerifItalics10 at 8pt
=\SansSerifSlanted10 at 8pt
=			\Math10				at 8pt
=		\MathBold10			at 8pt
=			\Symbols10			at 8pt
=		\MoreSymbols10		at 8pt
=		\Typewriter10		at 8pt
=			\Gothic10			at 8pt
=			\Double10			at 8pt

= 			\Serif10 			at 10pt
= 		\SerifBold10 		at 10pt
= 		\SerifItalics10 	at 10pt
=		\SerifSlanted10 	at 10pt
=	\SerifBoldItalics10	at 10pt
= 		\SansSerif10 		at 10pt
= 	\SansSerifBold10 	at 10pt
= 	\SansSerifItalics10 at 10pt
= 	\SansSerifSlanted10 at 10pt
=				\Math10				at 10pt
=			\MathBold10			at 10pt
=			\Symbols10			at 10pt
=		\MoreSymbols10		at 10pt
=		\Typewriter10		at 10pt
=			\Gothic10			at 10pt
=			\Double10			at 10pt

= 				\Serif10 			at 12pt
= 			\SerifBold10 		at 12pt
= 		\SerifItalics10 	at 12pt
=		\SerifSlanted10 	at 12pt
=	\SerifBoldItalics10	at 12pt
= 			\SansSerif10 		at 12pt
= 		\SansSerifBold10 	at 12pt
= 	\SansSerifItalics10 at 12pt
= 	\SansSerifSlanted10 at 12pt
=				\Math10				at 12pt
=			\MathBold10			at 12pt
=			\Symbols10			at 12pt
=		\MoreSymbols10		at 12pt
=			\Typewriter10		at 12pt
=				\Gothic10			at 12pt
=				\Double10			at 12pt

= 			\Serif10 			at 14pt
= 		\SerifBold10 		at 14pt
= 	\SerifItalics10 	at 14pt
=		\SerifSlanted10 	at 14pt
=	\SerifBoldItalics10	at 14pt
= 		\SansSerif10 		at 14pt
= 	\SansSerifBold10 	at 14pt
= \SansSerifSlanted10 at 14pt
= \SansSerifItalics10 at 14pt
=				\Math10				at 14pt
=			\MathBold10			at 14pt
=			\Symbols10			at 14pt
=		\MoreSymbols10		at 14pt
=		\Typewriter10		at 14pt
=			\Gothic10			at 14pt
=			\Double10			at 14pt

\def\NormalStyle{\parindent=5pt\parskip=3pt\normalbaselineskip=14pt%
\def\nt{\tenSerif}%
\def\rm{\fam0\tenSerif}%
\textfont0=\tenSerif\scriptfont0=\sevenSerif\scriptscriptfont0=\fiveSerif
\textfont1=\tenMath\scriptfont1=\sevenMath\scriptscriptfont1=\fiveMath
\textfont2=\tenSymbols\scriptfont2=\sevenSymbols\scriptscriptfont2=\fiveSymbols
\textfont3=\tenMoreSymbols\scriptfont3=\sevenMoreSymbols\scriptscriptfont3=\fiveMoreSymbols
\textfont\itfam=\tenSerifItalics\def\it{\fam\itfam\tenSerifItalics}%
\textfont\slfam=\tenSerifSlanted\def\sl{\fam\slfam\tenSerifSlanted}%
\textfont\ttfam=\tenTypewriter\def\tt{\fam\ttfam\tenTypewriter}%
\textfont\bffam=\tenSerifBold%
\def\bf{\fam\bffam\tenSerifBold}\scriptfont\bffam=\sevenSerifBold\scriptscriptfont\bffam=\fiveSerifBold%
\def\cal{\tenSymbols}%
\def\greekbold{\tenMathBold}%
\def\gothic{\tenGothic}%
\def\Bbb{\tenDouble}%
\def\LieFont{\tenSerifItalics}%
\nt\normalbaselines\baselineskip=14pt%
}

\def\TitleStyle{\parindent=0pt\parskip=0pt\normalbaselineskip=15pt%
\def\nt{\fourteenSansSerifBold}%
\def\rm{\fam0\fourteenSansSerifBold}%
\textfont0=\fourteenSansSerifBold\scriptfont0=\tenSansSerifBold\scriptscriptfont0=\eightSansSerifBold
\textfont1=\fourteenMath\scriptfont1=\tenMath\scriptscriptfont1=\eightMath
\textfont2=\fourteenSymbols\scriptfont2=\tenSymbols\scriptscriptfont2=\eightSymbols
\textfont3=\fourteenMoreSymbols\scriptfont3=\tenMoreSymbols\scriptscriptfont3=\eightMoreSymbols
\textfont\itfam=\fourteenSansSerifItalics\def\it{\fam\itfam\fourteenSansSerifItalics}%
\textfont\slfam=\fourteenSansSerifSlanted\def\sl{\fam\slfam\fourteenSerifSansSlanted}%
\textfont\ttfam=\fourteenTypewriter\def\tt{\fam\ttfam\fourteenTypewriter}%
\textfont\bffam=\fourteenSansSerif%
\def\bf{\fam\bffam\fourteenSansSerif}\scriptfont\bffam=\tenSansSerif\scriptscriptfont\bffam=\eightSansSerif%
\def\cal{\fourteenSymbols}%
\def\greekbold{\fourteenMathBold}%
\def\gothic{\fourteenGothic}%
\def\Bbb{\fourteenDouble}%
\def\LieFont{\fourteenSerifItalics}%
\nt\normalbaselines\baselineskip=15pt%
}

\def\PartStyle{\parindent=0pt\parskip=0pt\normalbaselineskip=15pt%
\def\nt{\fourteenSansSerifBold}%
\def\rm{\fam0\fourteenSansSerifBold}%
\textfont0=\fourteenSansSerifBold\scriptfont0=\tenSansSerifBold\scriptscriptfont0=\eightSansSerifBold
\textfont1=\fourteenMath\scriptfont1=\tenMath\scriptscriptfont1=\eightMath
\textfont2=\fourteenSymbols\scriptfont2=\tenSymbols\scriptscriptfont2=\eightSymbols
\textfont3=\fourteenMoreSymbols\scriptfont3=\tenMoreSymbols\scriptscriptfont3=\eightMoreSymbols
\textfont\itfam=\fourteenSansSerifItalics\def\it{\fam\itfam\fourteenSansSerifItalics}%
\textfont\slfam=\fourteenSansSerifSlanted\def\sl{\fam\slfam\fourteenSerifSansSlanted}%
\textfont\ttfam=\fourteenTypewriter\def\tt{\fam\ttfam\fourteenTypewriter}%
\textfont\bffam=\fourteenSansSerif%
\def\bf{\fam\bffam\fourteenSansSerif}\scriptfont\bffam=\tenSansSerif\scriptscriptfont\bffam=\eightSansSerif%
\def\cal{\fourteenSymbols}%
\def\greekbold{\fourteenMathBold}%
\def\gothic{\fourteenGothic}%
\def\Bbb{\fourteenDouble}%
\def\LieFont{\fourteenSerifItalics}%
\nt\normalbaselines\baselineskip=15pt%
}

\def\ChapterStyle{\parindent=0pt\parskip=0pt\normalbaselineskip=15pt%
\def\nt{\fourteenSansSerifBold}%
\def\rm{\fam0\fourteenSansSerifBold}%
\textfont0=\fourteenSansSerifBold\scriptfont0=\tenSansSerifBold\scriptscriptfont0=\eightSansSerifBold
\textfont1=\fourteenMath\scriptfont1=\tenMath\scriptscriptfont1=\eightMath
\textfont2=\fourteenSymbols\scriptfont2=\tenSymbols\scriptscriptfont2=\eightSymbols
\textfont3=\fourteenMoreSymbols\scriptfont3=\tenMoreSymbols\scriptscriptfont3=\eightMoreSymbols
\textfont\itfam=\fourteenSansSerifItalics\def\it{\fam\itfam\fourteenSansSerifItalics}%
\textfont\slfam=\fourteenSansSerifSlanted\def\sl{\fam\slfam\fourteenSerifSansSlanted}%
\textfont\ttfam=\fourteenTypewriter\def\tt{\fam\ttfam\fourteenTypewriter}%
\textfont\bffam=\fourteenSansSerif%
\def\bf{\fam\bffam\fourteenSansSerif}\scriptfont\bffam=\tenSansSerif\scriptscriptfont\bffam=\eightSansSerif%
\def\cal{\fourteenSymbols}%
\def\greekbold{\fourteenMathBold}%
\def\gothic{\fourteenGothic}%
\def\Bbb{\fourteenDouble}%
\def\LieFont{\fourteenSerifItalics}%
\nt\normalbaselines\baselineskip=15pt%
}

\def\SectionStyle{\parindent=0pt\parskip=0pt\normalbaselineskip=13pt%
\def\nt{\twelveSansSerifBold}%
\def\rm{\fam0\twelveSansSerifBold}%
\textfont0=\twelveSansSerifBold\scriptfont0=\eightSansSerifBold\scriptscriptfont0=\eightSansSerifBold
\textfont1=\twelveMath\scriptfont1=\eightMath\scriptscriptfont1=\eightMath
\textfont2=\twelveSymbols\scriptfont2=\eightSymbols\scriptscriptfont2=\eightSymbols
\textfont3=\twelveMoreSymbols\scriptfont3=\eightMoreSymbols\scriptscriptfont3=\eightMoreSymbols
\textfont\itfam=\twelveSansSerifItalics\def\it{\fam\itfam\twelveSansSerifItalics}%
\textfont\slfam=\twelveSansSerifSlanted\def\sl{\fam\slfam\twelveSerifSansSlanted}%
\textfont\ttfam=\twelveTypewriter\def\tt{\fam\ttfam\twelveTypewriter}%
\textfont\bffam=\twelveSansSerif%
\def\bf{\fam\bffam\twelveSansSerif}\scriptfont\bffam=\eightSansSerif\scriptscriptfont\bffam=\eightSansSerif%
\def\cal{\twelveSymbols}%
\def\bg{\twelveMathBold}%
\def\gothic{\twelveGothic}%
\def\Bbb{\twelveDouble}%
\def\LieFont{\twelveSerifItalics}%
\nt\normalbaselines\baselineskip=13pt%
}

\def\SubSectionStyle{\parindent=0pt\parskip=0pt\normalbaselineskip=13pt%
\def\nt{\twelveSansSerifItalics}%
\def\rm{\fam0\twelveSansSerifItalics}%
\textfont0=\twelveSansSerifItalics\scriptfont0=\eightSansSerifItalics\scriptscriptfont0=\eightSansSerifItalics%
\textfont1=\twelveMath\scriptfont1=\eightMath\scriptscriptfont1=\eightMath%
\textfont2=\twelveSymbols\scriptfont2=\eightSymbols\scriptscriptfont2=\eightSymbols%
\textfont3=\twelveMoreSymbols\scriptfont3=\eightMoreSymbols\scriptscriptfont3=\eightMoreSymbols%
\textfont\itfam=\twelveSansSerif\def\it{\fam\itfam\twelveSansSerif}%
\textfont\slfam=\twelveSansSerifSlanted\def\sl{\fam\slfam\twelveSerifSansSlanted}%
\textfont\ttfam=\twelveTypewriter\def\tt{\fam\ttfam\twelveTypewriter}%
\textfont\bffam=\twelveSansSerifBold%
\def\bf{\fam\bffam\twelveSansSerifBold}\scriptfont\bffam=\eightSansSerifBold\scriptscriptfont\bffam=\eightSansSerifBold%
\def\cal{\twelveSymbols}%
\def\greekbold{\twelveMathBold}%
\def\gothic{\twelveGothic}%
\def\Bbb{\twelveDouble}%
\def\LieFont{\twelveSerifItalics}%
\nt\normalbaselines\baselineskip=13pt%
}

\def\AuthorStyle{\parindent=0pt\parskip=0pt\normalbaselineskip=14pt%
\def\nt{\tenSerif}%
\def\rm{\fam0\tenSerif}%
\textfont0=\tenSerif\scriptfont0=\sevenSerif\scriptscriptfont0=\fiveSerif
\textfont1=\tenMath\scriptfont1=\sevenMath\scriptscriptfont1=\fiveMath
\textfont2=\tenSymbols\scriptfont2=\sevenSymbols\scriptscriptfont2=\fiveSymbols
\textfont3=\tenMoreSymbols\scriptfont3=\sevenMoreSymbols\scriptscriptfont3=\fiveMoreSymbols
\textfont\itfam=\tenSerifItalics\def\it{\fam\itfam\tenSerifItalics}%
\textfont\slfam=\tenSerifSlanted\def\sl{\fam\slfam\tenSerifSlanted}%
\textfont\ttfam=\tenTypewriter\def\tt{\fam\ttfam\tenTypewriter}%
\textfont\bffam=\tenSerifBold%
\def\bf{\fam\bffam\tenSerifBold}\scriptfont\bffam=\sevenSerifBold\scriptscriptfont\bffam=\fiveSerifBold%
\def\cal{\tenSymbols}%
\def\greekbold{\tenMathBold}%
\def\gothic{\tenGothic}%
\def\Bbb{\tenDouble}%
\def\LieFont{\tenSerifItalics}%
\nt\normalbaselines\baselineskip=14pt%
}

\def\AddressStyle{\parindent=0pt\parskip=0pt\normalbaselineskip=14pt%
\def\nt{\eightSerif}%
\def\rm{\fam0\eightSerif}%
\textfont0=\eightSerif\scriptfont0=\sevenSerif\scriptscriptfont0=\fiveSerif
\textfont1=\eightMath\scriptfont1=\sevenMath\scriptscriptfont1=\fiveMath
\textfont2=\eightSymbols\scriptfont2=\sevenSymbols\scriptscriptfont2=\fiveSymbols
\textfont3=\eightMoreSymbols\scriptfont3=\sevenMoreSymbols\scriptscriptfont3=\fiveMoreSymbols
\textfont\itfam=\eightSerifItalics\def\it{\fam\itfam\eightSerifItalics}%
\textfont\slfam=\eightSerifSlanted\def\sl{\fam\slfam\eightSerifSlanted}%
\textfont\ttfam=\eightTypewriter\def\tt{\fam\ttfam\eightTypewriter}%
\textfont\bffam=\eightSerifBold%
\def\bf{\fam\bffam\eightSerifBold}\scriptfont\bffam=\sevenSerifBold\scriptscriptfont\bffam=\fiveSerifBold%
\def\cal{\eightSymbols}%
\def\greekbold{\eightMathBold}%
\def\gothic{\eightGothic}%
\def\Bbb{\eightDouble}%
\def\LieFont{\eightSerifItalics}%
\nt\normalbaselines\baselineskip=14pt%
}

\def\AbstractStyle{\parindent=0pt\parskip=0pt\normalbaselineskip=12pt%
\def\nt{\eightSerif}%
\def\rm{\fam0\eightSerif}%
\textfont0=\eightSerif\scriptfont0=\sevenSerif\scriptscriptfont0=\fiveSerif
\textfont1=\eightMath\scriptfont1=\sevenMath\scriptscriptfont1=\fiveMath
\textfont2=\eightSymbols\scriptfont2=\sevenSymbols\scriptscriptfont2=\fiveSymbols
\textfont3=\eightMoreSymbols\scriptfont3=\sevenMoreSymbols\scriptscriptfont3=\fiveMoreSymbols
\textfont\itfam=\eightSerifItalics\def\it{\fam\itfam\eightSerifItalics}%
\textfont\slfam=\eightSerifSlanted\def\sl{\fam\slfam\eightSerifSlanted}%
\textfont\ttfam=\eightTypewriter\def\tt{\fam\ttfam\eightTypewriter}%
\textfont\bffam=\eightSerifBold%
\def\bf{\fam\bffam\eightSerifBold}\scriptfont\bffam=\sevenSerifBold\scriptscriptfont\bffam=\fiveSerifBold%
\def\cal{\eightSymbols}%
\def\greekbold{\eightMathBold}%
\def\gothic{\eightGothic}%
\def\Bbb{\eightDouble}%
\def\LieFont{\eightSerifItalics}%
\nt\normalbaselines\baselineskip=12pt%
}

\def\RefsStyle{\parindent=0pt\parskip=0pt%
\def\nt{\eightSerif}%
\def\rm{\fam0\eightSerif}%
\textfont0=\eightSerif\scriptfont0=\sevenSerif\scriptscriptfont0=\fiveSerif
\textfont1=\eightMath\scriptfont1=\sevenMath\scriptscriptfont1=\fiveMath
\textfont2=\eightSymbols\scriptfont2=\sevenSymbols\scriptscriptfont2=\fiveSymbols
\textfont3=\eightMoreSymbols\scriptfont3=\sevenMoreSymbols\scriptscriptfont3=\fiveMoreSymbols
\textfont\itfam=\eightSerifItalics\def\it{\fam\itfam\eightSerifItalics}%
\textfont\slfam=\eightSerifSlanted\def\sl{\fam\slfam\eightSerifSlanted}%
\textfont\ttfam=\eightTypewriter\def\tt{\fam\ttfam\eightTypewriter}%
\textfont\bffam=\eightSerifBold%
\def\bf{\fam\bffam\eightSerifBold}\scriptfont\bffam=\sevenSerifBold\scriptscriptfont\bffam=\fiveSerifBold%
\def\cal{\eightSymbols}%
\def\greekbold{\eightMathBold}%
\def\gothic{\eightGothic}%
\def\Bbb{\eightDouble}%
\def\LieFont{\eightSerifItalics}%
\nt\normalbaselines\baselineskip=10pt%
}

\def\ClaimStyle{\parindent=5pt\parskip=3pt\normalbaselineskip=14pt%
\def\nt{\tenSerifSlanted}%
\def\rm{\fam0\tenSerifSlanted}%
\textfont0=\tenSerifSlanted\scriptfont0=\sevenSerifSlanted\scriptscriptfont0=\fiveSerifSlanted
\textfont1=\tenMath\scriptfont1=\sevenMath\scriptscriptfont1=\fiveMath
\textfont2=\tenSymbols\scriptfont2=\sevenSymbols\scriptscriptfont2=\fiveSymbols
\textfont3=\tenMoreSymbols\scriptfont3=\sevenMoreSymbols\scriptscriptfont3=\fiveMoreSymbols
\textfont\itfam=\tenSerifItalics\def\it{\fam\itfam\tenSerifItalics}%
\textfont\slfam=\tenSerif\def\sl{\fam\slfam\tenSerif}%
\textfont\ttfam=\tenTypewriter\def\tt{\fam\ttfam\tenTypewriter}%
\textfont\bffam=\tenSerifBold%
\def\bf{\fam\bffam\tenSerifBold}\scriptfont\bffam=\sevenSerifBold\scriptscriptfont\bffam=\fiveSerifBold%
\def\cal{\tenSymbols}%
\def\greekbold{\tenMathBold}%
\def\gothic{\tenGothic}%
\def\Bbb{\tenDouble}%
\def\LieFont{\tenSerifItalics}%
\nt\normalbaselines\baselineskip=14pt%
}

\def\ProofStyle{\parindent=5pt\parskip=3pt\normalbaselineskip=14pt%
\def\nt{\tenSerifSlanted}%
\def\rm{\fam0\tenSerifSlanted}%
\textfont0=\tenSerif\scriptfont0=\sevenSerif\scriptscriptfont0=\fiveSerif
\textfont1=\tenMath\scriptfont1=\sevenMath\scriptscriptfont1=\fiveMath
\textfont2=\tenSymbols\scriptfont2=\sevenSymbols\scriptscriptfont2=\fiveSymbols
\textfont3=\tenMoreSymbols\scriptfont3=\sevenMoreSymbols\scriptscriptfont3=\fiveMoreSymbols
\textfont\itfam=\tenSerifItalics\def\it{\fam\itfam\tenSerifItalics}%
\textfont\slfam=\tenSerif\def\sl{\fam\slfam\tenSerif}%
\textfont\ttfam=\tenTypewriter\def\tt{\fam\ttfam\tenTypewriter}%
\textfont\bffam=\tenSerifBold%
\def\bf{\fam\bffam\tenSerifBold}\scriptfont\bffam=\sevenSerifBold\scriptscriptfont\bffam=\fiveSerifBold%
\def\cal{\tenSymbols}%
\def\greekbold{\tenMathBold}%
\def\gothic{\tenGothic}%
\def\Bbb{\tenDouble}%
\def\LieFont{\tenSerifItalics}%
\nt\normalbaselines\baselineskip=14pt%
}

%
%


\def\ModeYes{yes}
\def\ModeNo{no}

\def\ModeUndef{undefined}


\def\nx{\noexpand}
\def\ni{\noindent}
\def\newpage{\vfill\eject}

\def\ss{\vskip 5pt}
\def\ms{\vskip 10pt}
\def\bs{\vskip 20pt}

 \def\,{\mskip\thinmuskip}
 \def\!{\mskip-\thinmuskip}
 \def\>{\mskip\medmuskip}
 \def\;{\mskip\thickmuskip}

%
% ... 
%

\def\refsModePost{post}
\def\refsModeAuto{auto}

\def\dbRefsSatusModeOk{ok}
\def\dbRefsSatusModeError{error}
\def\dbRefsSatusModeWarning{warning}


\newcount\BNUM
\BNUM=0

\def\refs{}

\def\SetModePost{\xdef\refsMode{\refsModePost}}			
\SetModePost

\def\dbRefsStatusOk{%
	\xdef\dbRefsStatus{\dbRefsSatusModeOk}%
	\xdef\dbRefsError{\ModeNo}%
	\xdef\dbRefsWarning{\ModeNo}%
	\xdef\dbRefsInfo{\ModeNo}%
}

\def\dbRefs{%
}

\def\dbRefsGet#1{%
	\xdef\found{N}\xdef\ikey{#1}\dbRefsStatusOk%
	\xdef\key{\ModeUndef}\xdef\tag{\ModeUndef}\xdef\tail{\ModeUndef}%
	\dbRefs%
}

\def\NextRefsTag{%
	\global\advance\BNUM by 1%
}
\def\ShowTag#1{{\bf [#1]}}

\def\dbRefsInsert#1#2{%
\dbRefsGet{#1}%
\if\found Y %
   \xdef\dbRefsStatus{\dbRefsSatusModeWarning}%
   \xdef\dbRefsWarning{record is already there}%
   \xdef\dbRefsInfo{record not inserted}%
\else%
   \toks2=\expandafter{\dbRefs}%
   \ifx\refsMode\refsModeAuto \NextRefsTag
    \xdef\dbRefs{%
   	\the\toks2 \nx\xdef\nx\dbx{#1}%
	\nx\ifx\nx\ikey %
		\nx\dbx\nx\xdef\nx\found{Y}%
		\nx\xdef\nx\key{#1}%
		\nx\xdef\nx\tag{\the\BNUM}%
		\nx\xdef\nx\tail{#2}%
	\nx\fi}%
	\global\xdef\refs{\refs \ss\ni[\the\BNUM]\ #2\par}
   \fi%
   \ifx\refsMode\refsModePost 
    \xdef\dbRefs{%
   	\the\toks2 \nx\xdef\nx\dbx{#1}%
	\nx\ifx\nx\ikey %
		\nx\dbx\nx\xdef\nx\found{Y}%
		\nx\xdef\nx\key{#1}%
		\nx\xdef\nx\tag{\ModeUndef}%
		\nx\xdef\nx\tail{#2}%
	\nx\fi}%
   \fi%
\fi%
}

\def\dbRefsEdit#1#2#3{\dbRefsGet{#1}%
\if\found N 
   \xdef\dbRefsStatus{\dbRefsSatusModeError}%
   \xdef\dbRefsError{record is not there}%
   \xdef\dbRefsInfo{record not edited}%
\else%
   \toks2=\expandafter{\dbRefs}%
   \xdef\dbRefs{\the\toks2%
   \nx\xdef\nx\dbx{#1}%
   \nx\ifx\nx\ikey\nx\dbx %
	\nx\xdef\nx\found{Y}%
	\nx\xdef\nx\key{#1}%
	\nx\xdef\nx\tag{#2}%
	\nx\xdef\nx\tail{#3}%
   \nx\fi}%
\fi%
}

\def\bib#1#2{\RefsStyle\dbRefsInsert{#1}{#2}%
	\ifx\dbRefsStatus\dbRefsSatusModeWarning %
		\message{^^J}%
		\message{WARNING: Reference [#1] is doubled.^^J}%
	\fi%
}

\def\ref#1{\dbRefsGet{#1}%
\ifx\found N %
  \message{^^J}%
  \message{ERROR: Reference [#1] unknown.^^J}%
  \ShowTag{??}%
\else%
	\ifx\tag\ModeUndef \NextRefsTag%
		\dbRefsEdit{#1}{\the\BNUM}{\tail}%
		\dbRefsGet{#1}%
		\global\xdef\refs{\refs \ss\ni [\tag]\ \tail\par}
	\fi
	\ShowTag{\tag}%
\fi%
}

\def\ShowBiblio{\bs\Ensure{\SectionEnsure}%
{\SectionStyle\ni References}%
{\RefsStyle\refs}%
}

\newcount\CHANGES
\CHANGES=0
\def\AuxFile{7}
\def\PreventDoubleOn{\xdef\PreventDoubleLabel{\ModeYes}}

\PreventDoubleOn

\def\StoreLabel#1#2{\xdef\itag{#2}
 \ifx\PreModeStatus\ModeNo %
   \message{^^J}%
   \errmessage{You can't use Check without starting with OpenPreMode (and finishing with ClosePreMode)^^J}%
 \else%
   \immediate\write\AuxFile{\nx\dbLabelPreInsert{#1}{\itag}}%
   \dbLabelGet{#1}%
   \ifx\itag\tag %
   \else%
	\global\advance\CHANGES by 1%
 	\xdef\itag{(?.??)}%
    \fi%
   \fi%
}

\def\PreModeStatus{\ModeNo}

\def\ClosePreMode{\immediate\closeout\AuxFile%
  \ifnum\CHANGES=0%
	\message{^^J}%
	\message{**********************************^^J}%
	\message{**  NO CHANGES TO THE AuxFile  **^^J}%
	\message{**********************************^^J}%
 \else%
	\message{^^J}%
	\message{**************************************************^^J}%
	\message{**  PLAEASE TYPESET IT AGAIN (\the\CHANGES)  **^^J}%
    \errmessage{**************************************************^^ J}%
  \fi%
  \edef\PreModeStatus{\ModeNo}%
}

\def\dbLabelSatusModeOk{ok}

\def\dbLabelSatusModeWarning{warning}

\def\dbLabelStatusOk{%
	\xdef\dbLabelStatus{\dbLabelSatusModeOk}%
	\xdef\dbLabelError{\ModeNo}%
	\xdef\dbLabelWarning{\ModeNo}%
	\xdef\dbLabelInfo{\ModeNo}%
}

\def\dbLabel{%
}

\def\dbLabelGet#1{%
	\xdef\found{N}\xdef\ikey{#1}\dbLabelStatusOk%
	\xdef\key{\ModeUndef}\xdef\tag{\ModeUndef}\xdef\pre{\ModeUndef}%
	\dbLabel%
}

\def\ShowLabel#1{%
 \dbLabelGet{#1}%
 \ifx\tag \ModeUndef %
 	\global\advance\CHANGES by 1%
 	(?.??)%
 \else%
 	\tag%
 \fi%
}

\def\dbLabelPreInsert#1#2{\dbLabelGet{#1}%
\if\found Y %
  \xdef\dbLabelStatus{\dbLabelSatusModeWarning}%
   \xdef\dbLabelWarning{Label is already there}%
   \xdef\dbLabelInfo{Label not inserted}%
   \message{^^J}%
   \errmessage{Double pre definition of label [#1]^^J}%
\else%
   \toks2=\expandafter{\dbLabel}%
    \xdef\dbLabel{%
   	\the\toks2 \nx\xdef\nx\dbx{#1}%
	\nx\ifx\nx\ikey %
		\nx\dbx\nx\xdef\nx\found{Y}%
		\nx\xdef\nx\key{#1}%
		\nx\xdef\nx\tag{#2}%
		\nx\xdef\nx\pre{\ModeYes}%
	\nx\fi}%
\fi%
}

\def\dbLabelInsert#1#2{\dbLabelGet{#1}%
\xdef\itag{#2}%
\dbLabelGet{#1}%
\if\found Y %
	\ifx\tag\itag %
	\else%
	   \ifx\PreventDoubleLabel\ModeYes %
		\message{^^J}%
		\errmessage{Double definition of label [#1]^^J}%
	   \else%
		\message{^^J}%
		\message{Double definition of label [#1]^^J}%
	   \fi%
	\fi%
   \xdef\dbLabelStatus{\dbLabelSatusModeWarning}%
   \xdef\dbLabelWarning{Label is already there}%
   \xdef\dbLabelInfo{Label not inserted}%
\else%
   \toks2=\expandafter{\dbLabel}%
    \xdef\dbLabel{%
   	\the\toks2 \nx\xdef\nx\dbx{#1}%
	\nx\ifx\nx\ikey %
		\nx\dbx\nx\xdef\nx\found{Y}%
		\nx\xdef\nx\key{#1}%
		\nx\xdef\nx\tag{#2}%
		\nx\xdef\nx\pre{\ModeNo}%
	\nx\fi}%
\fi%
}


\newcount\PART
\newcount\CHAPTER
\newcount\SECTION
\newcount\SUBSECTION
\newcount\FNUMBER

\PART=0
\CHAPTER=0
\SECTION=0
\SUBSECTION=0	
\FNUMBER=0

\def\LastPart{\ModeUndef}
\def\LastChapter{\ModeUndef}
\def\LastSection{\ModeUndef}
\def\LastSubSection{\ModeUndef}
\def\LastClaim{\ModeUndef}
\def\Last{\ModeUndef}

\newdimen\TOBOTTOM
\newdimen\LIMIT

\def\Ensure#1{\ \par\ \immediate\LIMIT=#1\immediate\TOBOTTOM=\the\pagegoal\advance\TOBOTTOM by -\pagetotal%
\ifdim\TOBOTTOM<\LIMIT\newpage \else%
\vskip-\parskip\vskip-\parskip\vskip-\baselineskip\fi}

\def\PartLabel{\the\PART}
\def\NewPart#1{\global\advance\PART by 1%
         \bs\ni{\PartStyle  Part \PartLabel:}
         \bs\ni{\PartStyle #1}\newpage%
         \CHAPTER=0\SECTION=0\SUBSECTION=0\FNUMBER=0%
         \gdef\Left{#1}%
         \global\edef\Last{\PartLabel}%
         \global\edef\LastPart{\PartLabel}%
         \global\edef\LastChapter{\ModeUndef}%
         \global\edef\LastSection{\ModeUndef}%
         \global\edef\LastSubSection{\ModeUndef}%
         \global\edef\LastClaim{\ModeUndef}}
\def\ChapterLabel{\the\CHAPTER}
\def\NewChapter#1{\global\advance\CHAPTER by 1%
         \bs\ni{\ChapterStyle  Chapter \ChapterLabel: #1}\ms%
         \SECTION=0\SUBSECTION=0\FNUMBER=0%
         \gdef\Left{#1}%
         \global\edef\Last{\ChapterLabel}%
         \global\edef\LastChapter{\ChapterLabel}%
         \global\edef\LastSection{\ModeUndef}%
         \global\edef\LastSubSection{\ModeUndef}%
         \global\edef\LastClaim{\ModeUndef}}
\def\SectionEnsure{3cm}
\def\NewSection#1{\Ensure{\SectionEnsure}\gdef\SectionLabel{\the\SECTION}\global\advance\SECTION by 1%
         \bs\ni{\SectionStyle  \SectionLabel.\ #1}\ss%
         \SUBSECTION=0\FNUMBER=0%
         \gdef\Left{#1}%
         \global\edef\Last{\SectionLabel}%
         \global\edef\LastSection{\SectionLabel}%
         \global\edef\LastSubSection{\ModeUndef}%
         \global\edef\LastClaim{\ModeUndef}}
\def\NewAppendix#1#2{\Ensure{\SectionEnsure}\gdef\SectionLabel{#1}\global\advance\SECTION by 1%
         \bs\ni{\SectionStyle  Appendix \SectionLabel.\ #2}\ss%
         \SUBSECTION=0\FNUMBER=0%
         \gdef\Left{#2}%
         \global\edef\Last{\SectionLabel}%
         \global\edef\LastSection{\SectionLabel}%
         \global\edef\LastSubSection{\ModeUndef}%
         \global\edef\LastClaim{\ModeUndef}}
\def\Acknowledgements{\Ensure{\SectionEnsure}\gdef\SectionLabel{}%
         \bs\ni{\SectionStyle  Acknowledgments}\ss%
         \SECTION=0\SUBSECTION=0\FNUMBER=0%
         \gdef\Left{}%
         \global\edef\Last{\ModeUndef}%
         \global\edef\LastSection{\ModeUndef}%
         \global\edef\LastSubSection{\ModeUndef}%
         \global\edef\LastClaim{\ModeUndef}}
\def\SubSectionEnsure{2cm}
\def\SubSectionLabel{\ifnum\SECTION>0 \the\SECTION.\fi\the\SUBSECTION}
\def\NewSubSection#1{\Ensure{\SubSectionEnsure}\global\advance\SUBSECTION by 1%
         \ms\ni{\SubSectionStyle #1}\ss%
         \global\edef\Last{\SubSectionLabel}%
         \global\edef\LastSubSection{\SubSectionLabel}}
\def\SetNumberingModeN{\def\ClaimLabel{(\the\FNUMBER)}}
\def\SetNumberingModeSN{\def\ClaimLabel{(\ifnum\SECTION>0 \SectionLabel.\fi%
      \the\FNUMBER)}}
\def\SetNumberingModeCSN{\def\ClaimLabel{(\ifnum\CHAPTER>0 \the\CHAPTER.\fi%
      \ifnum\SECTION>0 \SectionLabel.\fi%
      \the\FNUMBER)}}

\def\NewClaim{\global\advance\FNUMBER by 1%
    \ClaimLabel%
    \global\edef\LastClaim{\ClaimLabel}%
    \global\edef\Last{\ClaimLabel}}

\def\HideLabels{\xdef\ShowLabelsMode{\ModeNo}}
\HideLabels

\def\fn{\eqno{\NewClaim}} 
\def\fl#1{%
\ifx\ShowLabelsMode\ModeYes%
 \eqno{{\buildrel{\hbox{\AbstractStyle[#1]}}\over{\hfill\NewClaim}}}%
\else%
 \eqno{\NewClaim}%
\fi%
\dbLabelInsert{#1}{\ClaimLabel}}
\def\fprel#1{\global\advance\FNUMBER by 1\StoreLabel{#1}{\ClaimLabel}%
\ifx\ShowLabelsMode\ModeYes%
\eqno{{\buildrel{\hbox{\AbstractStyle[#1]}}\over{\hfill.\itag}}}%
\else%
 \eqno{\itag}%
\fi%
}

\def\cl#1{\global\advance\FNUMBER by 1\dbLabelInsert{#1}{\ClaimLabel}%
\ifx\ShowLabelsMode\ModeYes%
${\buildrel{\hbox{\AbstractStyle[#1]}}\over{\hfill\ClaimLabel}}$%
\else%
  $\ClaimLabel$%
\fi%
}
\def\cprel#1{\global\advance\FNUMBER by 1\StoreLabel{#1}{\ClaimLabel}%
\ifx\ShowLabelsMode\ModeYes%
${\buildrel{\hbox{\AbstractStyle[#1]}}\over{\hfill.\itag}}$%
\else%
  $\itag$%
\fi%
}


\parindent=7pt
\leftskip=2cm
\newcount\SideIndent
\newcount\SideIndentTemp
\SideIndent=0
\newdimen\SectionIndent
\SectionIndent=-8pt

\def\sidebar{\vrule height15pt width.2pt }
\def\endcorner{\hbox{\hbox{\vrule height6pt width.2pt}\vbox to6pt{\vfill\hbox
to4pt{\leaders\hrule height0.2pt\hfill}}}}
\def\begincorner{\hbox{\hbox{\vrule height6pt width.2pt}\vbox to6pt{\hbox
to4pt{\leaders\hrule height0.2pt\hfill}}}}
\def\endbegincorner{\hbox{\vbox to15pt{\endcorner\vskip-6pt\begincorner\vfill}}}
\def\SideShow{\SideIndentTemp=\SideIndent \ifnum \SideIndentTemp>0 
\loop\sidebar\hskip 2pt \advance\SideIndentTemp by-1\ifnum \SideIndentTemp>1 \repeat\fi}

\def\BeginSection{{\vbadness 100000 \par\ni\hskip\SectionIndent%
\SideShow\vbox to 15pt{\vfill\begincorner}}\global\advance\SideIndent by1\vskip-10pt}

\def\EndSection{{\vbadness 100000 \par\ni\global\advance\SideIndent by-1%
\hskip\SectionIndent\SideShow\vbox to15pt{\endcorner\vfill}\vskip-10pt}}

\def\EndBeginSection{{\vbadness 100000\par\ni%
\global\advance\SideIndent by-1\hskip\SectionIndent\SideShow
\vbox to15pt{\vfill\endbegincorner}}%
\global\advance\SideIndent by1\vskip-10pt}

\def\ShowBeginCorners#1{%
\SideIndentTemp =#1 \advance\SideIndentTemp by-1%
\ifnum \SideIndentTemp>0 %
\vskip-15truept\hbox{\kern 2truept\vbox{\hbox{\begincorner}%
\ShowBeginCorners{\SideIndentTemp}\vskip-3truept}}%
\fi%
}

\def\ShowEndCorners#1{%
\SideIndentTemp =#1 \advance\SideIndentTemp by-1%
\ifnum \SideIndentTemp>0 %
\vskip-15truept\hbox{\kern 2truept\vbox{\hbox{\endcorner}%
\ShowEndCorners{\SideIndentTemp}\vskip 2truept}}%
\fi%
}

\def\BeginSections#1{{\vbadness 100000 \par\ni\hskip\SectionIndent%
\SideShow\vbox to 15pt{\vfill\ShowBeginCorners{#1}}}\global\advance\SideIndent by#1\vskip-10pt}

\def\EndSections#1{{\vbadness 100000 \par\ni\global\advance\SideIndent by-#1%
\hskip\SectionIndent\SideShow\vbox to15pt{\vskip15pt\ShowEndCorners{#1}\vfill}\vskip-10pt}}

\def\EndBeginSections#1#2{{\vbadness 100000\par\ni%
\global\advance\SideIndent by-#1%
\hbox{\hskip\SectionIndent\SideShow\kern-2pt%
\vbox to15pt{\vskip15pt\ShowEndCorners{#1}\vskip4pt\ShowBeginCorners{#2}}}}%
\global\advance\SideIndent by#2\vskip-10pt}




%
%


\def\al{\alpha}
\def\be{\beta}
\def\de{\delta}
\def\ga{\gamma}

\def\ep{\epsilon}

\def\te{\theta}
\def\la{\lambda}

\def\om{\omega}
\def\si{\sigma}
\def\vp{\varphi}

\def\ka{\kappa}

\def\De{\Delta}
\def\Ga{\Gamma}

\def\La{\Lambda}

\def\Si{\Sigma}


 \def\su{{\hbox{\gothic su}}}


 \def\one{{\hbox{\Bbb I}}}
 
 \def\R{{\hbox{\Bbb R}}}
 \def\C{{\hbox{\Bbb C}}}
 
 \def\E{{\hbox{\Bbb E}}}

 \def\R{{\hbox{\Bbb R}}}


\def\Spin{{\hbox{Spin}}}
\def\SO{{\hbox{SO}}}
\def\SU{{\hbox{SU}}}
\def\SL{{\hbox{SL}}}
\def\GL{{\hbox{GL}}}
\def\det{{\hbox{det}}}

\def\di{{\hbox{d}}}
\def\id{{\hbox{\rm id}}}

\def\ip{\hbox to4pt{\leaders\hrule height0.3pt\hfill}\vbox to8pt{\leaders\vrule width0.3pt\vfill}\kern 2pt}
\def\QDE{\hfill\hbox{\ }\vrule height4pt width4pt depth0pt} 
\def\del{\partial}
\def\na{\nabla}

\def\arr{\rightarrow}
\def\harr{\hookrightarrow}
\def\then{\Rightarrow}

%
%

\def\LEMMAl#1{\ClaimStyle\ni{\bf Lemma \cl{#1}: }}

\def\ENDLEMMA{\NormalStyle}

\def\PROOF{\ProofStyle\ni{\bf Proof: }}
\def\ENDPROOF{\hfill\QDE\NormalStyle}

\def\cases#1{\left\{\eqalign{#1}\right.}
\NormalStyle
\SetNumberingModeSN
\PreventDoubleOn

\long\def\title#1{\centerline{\TitleStyle\ni#1}}
\long\def\author#1{\ms\centerline{\AuthorStyle by {\it #1}}}

\long\def\address#1{\ss\centerline{\AddressStyle #1}\par}
\long\def\moreaddress#1{\centerline{\AddressStyle #1}\par}
\def\abstract{\ms\leftskip 3cm\rightskip .5cm\AbstractStyle{\bf \ni Abstract:}\ }
\def\endabstract{\par\leftskip 2cm\rightskip 0cm\NormalStyle\ss}

\SetNumberingModeSN

\def\nab#1{{\buildrel #1\over \na}}
\def\frac[#1/#2]{\hbox{$#1\over#2$}}

\def\({\left(}
\def\){\right)}
\def\[{\left[}
\def\]{\right]}
\def\^#1{{}^{#1}_{\>\cdot}}
\def\_#1{{}_{#1}^{\>\cdot}}
\def\Label=#1{{\buildrel {\hbox{\fiveSerif \ShowLabel{#1}}}\over =}}
\def\<{\kern -1pt}

\bib{OurAshtekar}{L.\ Fatibene, M.\ Francaviglia, {\it Spin Structures on Manifolds and Ashtekar Variables}, Int. J. Geom. Methods Mod. Phys. {\bf 2}(2), 147-157, (2005)}

\bib{OurBI}{L.\ Fatibene, M.\ Francaviglia, C.\ Rovelli, {\it On a Covariant Formulation of the Barbero-Immirzi Connection}, Class. Quantum Grav. 24 (2007) 3055-3066.}

\bib{Book}{L.\ Fatibene, M.\ Francaviglia, {\it Natural and gauge natural formalism for classical field theories. A geometric perspective including spinors and gauge theories}, Kluwer Academic Publishers, Dordrecht, 2003}

\bib{Perez}{A. Perez, {\it Introduction to Loop Quantum Gravity and Spin Foams},
in: Proceedings of the II International Conference on Fundamental Interactions, Pedra Azul, Brazil, June 2004; gr-qc/0409061}

\bib{Holst}{S.\ Holst, {\it Barbero's Hamiltonian Derived from a Generalized Hilbert-Palatini Action},
Phys.\ Rev.\ {\bf D53}, 5966, 1996}

\bib{RovelliBook}{C.\ Rovelli, {\it Quantum Gravity}, Cambridge University Press, (Cambridge, 2004)}

\bib{RovelliBI}{A.\ Perez, C.\ Rovelli, {\it Physical effects of the Immirzi parameter},
Phys.\ Rev.\ {\bf D73}, 044013, 2006}

\bib{gr-qc/0404018}{A.\ Ashtekar, J.\ Lewandowski,
{\it Background Independent Quantum Gravity: a Status Report}, gr-qc/0404018}

\bib{Thiemann}{T.\ Thiemann, {\it Introduction to Modern Canonical Quantum General Relativity};
gr-qc/0110034}

\bib{Barbero}{F.\ Barbero, {\it Real Ashtekar variables for Lorentzian signature space-time},
Phys.\ Rev.\ {\it D51}, 5507, (1996)}

\bib{Immirzi}{G.\ Immirzi, {\it Quantum Gravity and Regge Calculus},
Nucl.\ Phys.\ Proc.\ Suppl.\ {\bf 57}, 65-72}

\bib{KobaNu}{S.\ Kobayashi, K.\ Nomizu,
  {\it Foundations of differential geometry},
  John Wiley \& Sons, Inc., New York, 1963 USA}

\bib{Antonsen}{F.\ Antonsen, M.S.N.\ Flagga, {\it Spacetime Topology (I) - Chirality and the Third Stiefel-Whitney Class}, Int.\ J.\ Th.\ Phys.\ {\bf 41}(2), 2002}

\bib{Milnor}{J.W.\ Milnor, J.\ Stasheff, {\it Characteristic classes}, Princeton University Press, (Princeton NJ, USA, 1974)}

\NormalStyle
\NormalStyle
\def\FieldEqsEquivalenceAPP{A}
\def\FrameAPP{B}
\def \ProjectionIdentitiesAPP{C}

\title{Spacetime Lagrangian Formulation of Barbero-Immirzi Gravity}

\author{L. Fatibene$^{1, 2}$, M. Francaviglia$^{1, 2, 3}$, C. Rovelli$^{4}$}

\address{$^1$ Department of Mathematics, University of Torino (Italy)}

\moreaddress{$^2$ INFN- Iniziativa Specifica Na12}

\moreaddress{$^3$ ESG - University of Calabria (Italy)}

\moreaddress{${}^4$ Centre de Physique Th\'eorique de Luminy%
\footnote{$^\dagger$}{Unit\'e mixte de recherche (UMR 6207) du CNRS et des Universit\'es
de Provence (Aix-Marseille I), de la M\'editerran\'ee (Aix-Marseille
II) et du Sud (Toulon-Var); laboratoire affili\'e \`a la FRUMAM (FR
2291).},  Universit\'e de la M\'editerran\'ee, F-13288 Marseille, EU}

\abstract
We shall here discuss a new spacetime gauge-covariant Lagrangian formulation of General Relativity by means of the
Barbero-Immirzi $SU(2)$-connection on spacetime.
To the best of our knowledge the Lagrangian based on $\SU(2)$ spacetime fields seems to appear here for the first time.
\endabstract

\NewSection{Introduction}

In a previous paper of ours \ref{OurBI} we introduced new gauge-covariant spacetime variables that are suited to provide a spacetime interpretation of the Barbero-Immirzi $\SU(2)$-connection (BI connection) that, in turn, enters the formulation of Loop Quantum Gravity (LQG).

We shall here discuss the classical dynamics of General Relativity (GR) in terms of these new spacetime variables introduced in \ref{OurAshtekar} and \ref{OurBI}.

The main idea is very simple: we shall pull-back Holst's action in the new variables and then restrict it to
a spacelike hypersurface $S\subset M$ to obtain the constraint equations; 
see \ref{RovelliBook} for theoretical motivations.

The new Holst's Lagrangian is a functional of the spacetime fundamental fields 
$(e_a^\mu, A^i_\mu, K^i_\mu)$. 
These fields have to be considered as being independent. 
We shall call this Lagrangian the {\it Barbero-Immirzi-Holst (BIH) Lagrangian}.

Field equations of the BIH action will provide manifestly gauge-covariant equations that are in fact equivalent to the field equations of Holst action (though directly written in terms of the spacetime BI connection), which are in turn equivalent to standard GR equations.

By projecting onto a hypersurface $S\subset M$ field equations split into some evolution equations,
some constraint equations (that are in fact the starting point for LQG; see \ref{RovelliBook}, \ref{Perez}, \ref{Thiemann}) and
some further {\it (algebraic)} constraint equations which determine the $K$ field as a function of  the (densitized) triad $E_i^A$ and the BI connection $A^i_A$.

Although the general relation with the Hamiltonian multisymplectic framework will be investigated elsewhere it must be noticed here that the equations obtained by projection of the Lagrange equations on a hypersurface $S\subset M$ actually coincide for GR with the Hamiltonian constraints (see \ref{RovelliBook}).

This derivation is quite simple from the conceptual and computational viewpoint.
Moreover, it is quite useful to discuss on the Lagrangian side the gauge properties of the model together with its relations to other equivalent frameworks, such as the Dirac-Bergman Hamiltonian reduction.

\ 

\NewSection{Notation}

 We shall here briefly recall the notation introduced  \ref{OurBI}
 and adapt to the present case the results developed in \ref{OurAshtekar} for the selfdual case.

Let $M$ be an orientable connected and paracompact manifold of dimension $m=4$.
Let us fix either the Euclidean (or the Lorentzian) signature $\eta=(4,0)$ (or $\eta=(3,1)$, respectively).
For later convenience, we shall introduce a signature dependent quantity $\si$,
being $\si=1$ in the Euclidean  signature and  $\si=i$ in the Lorentzian one.

The spacetime manifold is assumed to allow global $\eta$-metrics 
and spin structures of the relevant signature.
This is equivalent to require the first and second Stiefel-Whitney classes of $M$ to vanish,
which in turn implies that the third Stiefel-Whitney class vanishes as well; see \ref{Antonsen} and \ref{Milnor}.

The group $\Spin(4)$ is known to be canonically isomorphic to $\SU(2)\times \SU(2)$.
The first $\SU(2)$ factor is called the selfdual part of the spin group,
while the second factor is the antiselfdual part;
the projection on the first factor is $p_+:\Spin(4)\arr \SU(2)$.

We also introduce a group homomorphism $\iota:\SU(2)\arr \Spin(\eta)$;
in the Euclidean case the group homomorphism $\iota:\SU(2)\arr \Spin(4)$ is defined as 
$\iota(S)=(S,S)$, where the isomorphism $\Spin(4)\simeq \SU(2)\times \SU(2)$ has been understood.
In the Lorentzian case the spin group $\Spin(3,1)$ is canonically isomorphic to $\SL(2,\C)$,
which is the ``complexified'' version of $\SU(2)$; in this case the group homomorphism
$\iota:\SU(2)\arr \Spin(3,1)$ exhibits $\SU(2)$ as a real section of $\SL(2, \C)$.

Let us choose a $\Spin(\eta)$-principal bundle $P$ over $M$ such that
global spin frames exist; see \ref{Book}.
We stress that this is considerably less than asking $M$ to be parallelizable (namely, to allow global sections of the general bundle $L(M)$ of frames, or, equivalently, the tangent bundle $TM$ to be trivial). For example, one can define this structure on all spheres despite the even dimensional spheres are not parallelizable.

The vanishing of the third Stiefel-Whitney class implies (see \ref{Antonsen} and \ref{Milnor}) the existence of a $\SU(2)$-reduction, namely of 
a $\SU(2)$-principal bundle $^+\<\<P$ together with a principal morphism relative to the 
group morphism $\iota:\SU(2)\arr \Spin(\eta)$
$$
\begindc{\commdiag}[1]
\obj(110,80)[+P2]{$^+\<\<P$}
\obj(180,80)[hP3]{$P$}
\obj(110,30)[M2]{$M$}
\obj(180,30)[M3]{$M$}
\mor{+P2}{M2}{}
\mor{hP3}{M3}{}
\mor{+P2}{hP3}{}
\mor{M2}{M3}{}[\atleft, \solidline] \mor(110,33)(180,33){}[\atleft, \solidline]
\enddc
\fl{ReductionDiag}$$

A local trivialization (also known as a {\it local gauge}) of $^+\<\<P$ amounts to fixing a local section
$\si^{(\al)}:U_\al\subset M \arr {}^+\<\<P$ on a chart domain $U_\al$.
Using the reduction \ShowLabel{ReductionDiag} a local trivialization of $^+\<\<P$ induces a local trivialization $\hat\si^{(\al)}$ of $P$.
By construction, two local trivializations on $^+\<\<P$ are mapped one into the other by a $\SU(2)$-gauge transformation $\si^{(\be)}= \si^{(\al)}\cdot \vp^{(\al\be)}(x)$ where $\vp^{(\al\be)}(x)\in \SU(2)$.
The same happens on $P$, namely, $\hat \si^{(\be)}=\hat \si^{(\al)}\cdot \iota\(\vp^{(\al\be)}(x)\)$
Hence $P$ allows, by construction, a trivialization with transition functions with values in $\SU(2)\harr\Spin(\eta)$.
For future convenience, we shall use on $P$ only this sort of reduced trivializations.

The standard Holst's fields are a spin tetrad $e_a^\mu$ and a spin connection $\te^{ab}_\mu$.
The spin frame $e_a^\mu$ is a global section of the bundle $P_\la$ associated to
$P\times L(M)$ by means of the following action
of the group $\GL(4)\times \Spin(\eta)$ on $\GL(4)$ (which has here to be considered as a manifold)
$$
\la:\GL(4)\times \Spin(\eta) \times \GL(4)\arr\GL(4):(J, S, e)\mapsto J\cdot e \cdot\ell(S^{-1})
\fn$$
where $\ell:\Spin(\eta)\arr \SO(\eta)$ is the covering map exhibiting the spin group as a double covering of the relevant orthogonal group.
The spin frame bundle $P_\la$ has local coordinates $(x^\mu, e_a^\mu)$ and, by construction, it is assumed to allow global sections (also when $M$ is non-parallelizable).
This framework is equivalent to the one dealing with soldering forms; see \ref{OurAshtekar} or \ref{Book} for details.

We can then replace the spin connection $\te^{ab}_\mu$ with the new variables
$$
\cases{
&A^i_\mu= \frac[1/2]\ep^i{}_{jk}\> \te^{jk}_\mu +\ga\te^{0i}_\mu\cr
&K^i_\mu= \te^{0i}_\mu\cr
}
\qquad\qquad\ga\in \R-\{0\}
\fl{BIVariables}$$
As discussed in \ref{OurBI}, $A^i_\mu$ is a $\SU(2)$-connection on ${}^+\<\<P$ while 
$K^i_\mu$ is a $\su(2)$-valued $1$-form on ${}^+\<\<P$.
We shall call $A^i_\mu$  the {\it (spacetime) BI connection} and $K^i_\mu$ the {\it extrinsic (spacetime) field}. Both fields live on a bundle associated to ${}^+\<\<P$ and are hence $\SU(2)$-objects.

\NewSection{BIH Lagrangian}

The standard Holst's Lagrangian (see \ref{Holst}, \ref{gr-qc/0404018}) reads as
$$
L_\ga(j^1\te, e) = \frac[1/4\ka]R^{ab} \land e^c\land e^d \ep_{abcd} 
+\frac[1/2\ka\ga]R^{ab} \land e_a \land e_b
\fl{HolstLagrangian}$$
where $R^{ab}$ denotes the curvature $2$-form of the spin connection $\te^{ab}_\mu$.
Let us denote by $\na$ the covariant derivative with respect to the connection $A^i$.

By using the transformation \ShowLabel{BIVariables} it is easy to prove that:
$$
\cases{
&R^{0i}=2\( \na K^i + \ga \ep^i{}_{jk}K^j\land K^k\)
=2\tilde \na K^i \cr
&R^{ij}= \ep^{ij}{}_k F^k -2\ga\ep^{ij}{}_k  \na K^k -2(\si^2+\ga^2) K^i\land K^j\cr
}
\fl{CurvatureIdentities}$$
where $\tilde\na$ denotes the covariant derivative with respect to the connection $A-\ga K$.

We can now use  \ShowLabel{CurvatureIdentities} to pull-back the Lagrangian
\ShowLabel{HolstLagrangian} along the new variables \ShowLabel{BIVariables} to obtain
$$
\eqalign{
{\bf L}_\ga(e, j^1A, j^1 K) =&
-\frac[1/4\ka\ga] \Big(
8(\si^2-\ga^2) \na K^i\land e_i\land e^0 
+4\ga(\si^2-\ga^2)\ep_{ijk} K^i\land K^j\land e^k\land e^0 +\cr
&+4\ga F^k\land e_k \land e^0
-2\ep_{ijk} F^i\land e^j\land e^k
-4(\si^2-\ga^2)K^i\land e_i\land K^j\land e_j
\Big)
}
\fl{HBILagrangian}$$
(here $j^1$ refers to the first order jet prolongations, simply meaning that the Lagrangian depends on the fields $(A^i_\mu, K^i_\mu)$ together with their first derivatives).
We stress that at this level the fields $(e_a^\mu, A^i_\mu, K^i_\mu)$ have to be considered
as independent fields.
By varying the Lagrangian \ShowLabel{HBILagrangian} we obtain field equations under the form
$$
\cases{
& 
\di e^0 \land e_i 
+K^j \land e_j\land e_i
=\ga\ep_{ijk} K^j\land e^k\land e^0
+ \na e_i  \land e^0
\cr
&  
\ga\di e^0\land e_k  
+ \ep_{ijk} \na e^i\land e^j
=\ga \na e_k \land  e^0
-(\si^2-\ga^2) \ep^i{}_{jk} K^i\land e_i\land e^0
\cr
&
F^k\land e_k
+(\si^2-\ga^2) \ep_{ijk} K^i\land K^j\land e^k
+2\frac[\si^2-\ga^2/\ga]\na K^i\land e_i
=0
\cr
&
2(\si^2-\ga^2) K^i\land e_i\land K^k
+\ep^k{}_{ij} F^i\land e^j
+(\si^2-\ga^2)\na K^k\land e^0 +
\cr
&\qquad
+\ga(\si^2-\ga^2) \ep^k{}_{ij} K^i\land K^j\land e^0
+\ga F^k\land e^0 
=0
\cr
}
\fl{BIHFieldsEquations}$$
These field equations are obtained by varying the Lagrangian  ${\bf L}_\ga$ with respect to the fields 
$(K^i_\mu,  A^i_\mu, e_0^\mu, e_k^\mu)$, respectively.

One can easily check that these field equations are equivalent to the field equations of the standard Holst's Lagrangian \ShowLabel{HolstLagrangian}, as expected; see Appendix $\FieldEqsEquivalenceAPP$.

\NewSection{Hamiltonian Framework}

Let us now fix a (spacelike) hypersurface $S\subset M$. We stress that we are fixing a single hypersurface, not a foliation.
Let us choose coordinates $k^A$ on $S$ so that the canonical injection is locally expressed
by $i:S\arr M:k\mapsto x(k)$.

The structure bundles \ShowLabel{ReductionDiag} can be pulled-back (i.e.~restricted) to $S$
obtaining:
$$
\begindc{\commdiag}[1]
\obj(110,80)[+P2]{$^+\<\<P$}
\obj(180,80)[hP3]{$P$}
\obj(70,50)[+Si4]{$^+\Si$}
\obj(140,50)[hSi5]{$\Si$}
\obj(110,30)[M2]{$M$}
\obj(180,30)[M3]{$M$}
\obj(70,0)[S4]{$S$}
\obj(140,0)[S5]{$S$}
\mor{+P2}{M2}{}
\mor{hP3}{M3}{}
\mor{+Si4}{S4}{}
\mor{hSi5}{S5}{}
\mor{+P2}{hP3}{}
\mor{M2}{M3}{}[\atleft, \solidline] \mor(110,33)(180,33){}[\atleft, \solidline]
\mor{S4}{S5}{}[\atleft, \solidline] \mor(70,3)(140,3){}[\atleft, \solidline]
\mor{S4}{M2}{}[\atleft, \injectionarrow]
\mor{S5}{M3}{}[\atleft, \injectionarrow]
\mor{+Si4}{+P2}{}[\atleft, \injectionarrow]
\mor{hSi5}{hP3}{}[\atleft, \injectionarrow]
\mor{+Si4}{hSi5}{}
\enddc
\fl{HamiltonianReductionDiag}$$
The bundles $^+\<\Si$ and $\Si$ are $\SU(2)$ and $\Spin(\eta)$ bundles, respectively.

As shown in \ref{OurAshtekar} by techniques adapted to the case of BI connection (see Appendix $\FrameAPP$) the spin tetrad $e_a^\mu$ canonically determines a spin triad $\ep_i^A$
on $S$  together with a vector $u$ normal to $S$.
Of course one has the identities
$$
u_\mu u^\mu=\si^2
\qquad
u_\mu e_i^\mu=0
\qquad
u_\mu \del_a x^\mu=0
\fn$$
that express the fact that $u$ is orthogonal to $S$ and it is unitary (and timelike in the Lorentzian case) with respect to the metric induced by the frame itself.

Notice that the frame $e_a^\mu$ is expressed by $4\times 4 =16$ functions.
The spin triad $\ep_i^A$ is expressed by $3\times 3=9$ functions, $u$ is expressed by $4$ functions.
As shown in  \ref{OurAshtekar},  the reduction on $S$ is achieved by canonically determining (out of the frame and the hypersurface $S$) an antiselfdual transformation, which is expressed by other $3=\dim\SU(2)$ functions. Thus the new variables are again described by $9+4+3=16$ functions (though $3$ of them---namely, the ones connected to the antiselfdual transformation--- are canonically fixed as a functions of the others once $S$ is fixed; for this reason they will be systematically dropped below).

The BI connection $A^i_\mu$ induces two fields on $S$, namely
$$
\cases{
& A^i_A= A^i_\mu \del_Ax^\mu\cr
&\tilde A^i= A^i_\mu u^\mu\cr
}
\fn$$ 
Similarly, the extrinsic field induces
$$
\cases{
& K^i_A= K^i_\mu \del_Ax^\mu\cr
&\tilde K^i= A^i_\mu u^\mu\cr
}
\fn$$
The field $A^i_A$ is a $\SU(2)$-connection on ${}^+\<\Si$ while $K^i_A$ is a $\su(2)$-valued $1$-form
on ${}^+\<\Si$.
The fields $(\tilde A^i, \tilde K^i)$ are $\su(2)$-valued scalar fields on ${}^+\<\Si$.

Notice that the original spin connection $\te^{ab}_\mu$ is expressed by $6\times 4=24$ functions.
Similarly, $A^i_\mu$ and $K^i_\mu$ are in fact $3\times 4=12$ functions each.
Once projected on $S$ we still have $3\times 3=9$ functions for $A^i_A$,
$3$ functions for $\tilde A^i$, $3\times 3=9$ functions for $K^i_A$,
$3$ functions for $\tilde K^i$, thus again $9+3+9+3=24$ functions.
Preserving the number of independent quantities is of course necessary for any change of variables in the space of fields.

We are now going to project the field equations \ShowLabel{BIHFieldsEquations}
onto $S$, i.e.~writing them in terms of the fields $(\ep_A^i, A^i_A, \tilde A^i, K^i_A, \tilde K^i)$;
see Appendix $\ProjectionIdentitiesAPP$ for technicalities.
We shall systematically split each equation into its component parallel to $S$
(by multiplying it by $\ep^{ABC}\del_Ax^\mu\del_Bx^\nu\del_Cx^\rho$)
and into its component orthogonal to  $S$
(by multiplying it by $\ep^{ABC} u^\mu\del_Bx^\nu\del_Cx^\rho$).

\

\NewSubSection{Equation for $\de K^i_\mu$}

Let us project the equation 
$$
\di e^0 \land e_i 
+K^j \land e_j\land e_i
=\ga\ep_{ijk} K^j\land e^k\land e^0
+ \na e_i  \land e^0
\fl{FirstFieldEquation}$$

The part parallel to $S$ is:
$$
\eqalign{
\ep^{ABC}\( \ep^i_{A} \di_B u_C + K^j_A \ep_{jB} \ep^i_{C}\)=0
}
\fn$$
Now using equation $(C.5)$ and multiplying by $\ep_{EFD}\ep^{D}_i$ we
obtain the equation
$$
K^j_{[A} \ep_{jB]} =0
\fl{KeShew}$$

We remark that this is a consequence of spacetime field equations of the model under consideration
and it is an {\it algebraic} relation between quantities defined on $S$.
We shall refer this sort of equations as {\it constraints}, in opposition to {\it evolution equations} which will appear later. 
Hereafter, we shall be particularly interested in constraint equations; see \ref{RovelliBook}.

The part orthogonal to $S$ is
$$
\eqalign{
\ep^i_{[A} \di_{B]} u_0 
+ \tilde K^j \ep_{j[A} \ep^i_{B]}
=&
\si^2 \na_{[A} \ep^i_{B]}
+\si^2\ga\ep^i{}_{jk} K^j_{[A}K^k_{B]}
\qquad \then\cr
\ep^{ABC}\big(2\ep^i_{A} \di_{[B} u_{0]} 
+ \tilde K^j \ep_{jA} \ep^i_{B}&
-\si^2 \nab{A-\ga K}_{A} \ep^i_{B}\big)
=0
\qquad \then\cr   
2\ep^{DBC}\di_{[B} u_{0]} 
+\ep^{ADC} \tilde K^j \ep_{jA}  &
-\si^2 \ep^{ABC}\ep_i^{D}  \>\nab{A-\ga K}_{A} \ep^i_{B}
=0
\cr
}
\fl{FirstOrthogonal}$$
We stress that we used above the expression $\di_{[B} u_{0]} $ in place of $\del_B x^\mu \di_{[\mu} e^0_{\nu]} u^\nu$.
As usual $[\cdot,\cdot ]$ denotes antisymmetrization of homologous indices.
Hereafter we shall set $\Ga=A-\ga K$.

Equations \ShowLabel{FirstOrthogonal} are $9$ equations.
Notice that the evolution part of these equations (namely, $2\ep^{DBC}\di_{[B} u_{0]}  $)
enters through the antisymmetric part in $[CD]$.
Let us thus split these $9$ equations into the symmetric ($6$ equations) and antisymmetric ($3$ equations) part.

The antisymmetric part is:
$$
2\ep^{ADC}\di_{[A} u_{0]} 
-\ep^{ADC} \tilde K^j \ep_{jA} 
+\si^2 \ep^{AB[C}_{\phantom{i}} \ep_i^{D]}  \>\nab{\Ga}_{A} \ep^i_{B}
=0
\fl{SkewFirstOrthogonal}$$

The symmetric part is:
$$
\ep^{AB(C}_{\phantom{i} }\ep_i^{D)}  \>\nab{\Ga}_{A} \ep^i_{B}
=0
\fl{SymmFirstOrthogonal}$$

Notice that these are $6$ further constraint equations, while equation \ShowLabel{SkewFirstOrthogonal} is not. 
We remark how the original $12$ equations  \ShowLabel{FirstFieldEquation}  have been split into $9$ constraint equations \ShowLabel{KeShew} and \ShowLabel{SymmFirstOrthogonal} and $3$ evolution equations  \ShowLabel{SkewFirstOrthogonal} .

\NewSubSection{Equation for $\de A^i_\mu$}
Let us project the equation 
$$
\ga\di e^0\land e_k  
+ \ep_{ijk} \na e^i\land e^j
=\ga \na e_k \land  e^0
-(\si^2-\ga^2) \ep^i{}_{jk} K^i\land e_i\land e^0
\fn$$

The part parallel to $S$ is
$$
\eqalign{
&\ep^{ABC}\(\ga \di_A u_B \ep^k_C+\ep^k{}_{ij} \na_A \ep^i_B \ep^j_C\)=0
\qquad\then\cr
&\ep^{ABC}\ep^k{}_{ij} \na_A \ep^i_B \ep^j_C
=\ep^{ABC}\ep \ep_{DEC} \ep^D_i\ep^E_j \na_A \ep^i_B
=2\ep \ep^{[A}_i\ep^{B]}_j \na_A \ep^i_B
=0
\qquad\then\cr
&\ep \ep^{A}_i\ep^{B}_j \na_A \ep^i_B
-\ep \ep^{B}_i\ep^{A}_j \na_A \ep^i_B
=0
\quad\then\quad
\ep \na_A\ep^{A}_i  \ep^i_C
+\ep \ep^{A}_i \na_C \ep^i_A
=0
\quad\then\quad \cr
&
\ep \na_A\ep^{A}_i  \ep^i_C
+ \na_C \ep
=0
\quad\then\quad
\ep \na_A\ep^{A}_i 
+ (\na_A \ep) \ep_i^A
=0
\quad\then\quad 
\na_A E^A_i=0
}
\fn$$
where we set $\ep:=\det \ep^i_A$ and $E_i^A:=\ep \ep_i^A$.

This constraint condition can be also expressed in the following form:
$$
\eqalign{
&\ep^{ABC}\na_{A} \ep^{[i}_B \ep^{j]}_{C}=0
\quad\then\quad
\ep^{AB[C}_{\phantom{i}} \ep_{i}^{D]} \na_{A} \ep^{i}_B =0
\quad\then\quad\cr
&
\ep^{AB[C}_{\phantom{i}} \ep_{i}^{D]} \nab{\Ga}_{A} \ep^{i}_B = 
\ga \ep^{AB[C}_{\phantom{i}} \ep_{i}^{D]} \ep^i{}_{jk}K^j_A\ep^{k}_B =
-\ga \ep K^i_{[C}\ep^{\phantom{i}}_{iD]} =0
}
\fl{SecondSymmEQ}$$
where we used \ShowLabel{KeShew} in the last equation.

Notice that \ShowLabel{SymmFirstOrthogonal} and \ShowLabel{SecondSymmEQ} together imply
$$
\ep^{ABC}_{\phantom{i}} \ep_{i}^{D} \nab{\Ga}_{A} \ep^{i}_B=0
\quad\then\quad
\nab{\Ga}_{[A} \ep^{i}_{B]}=0
\fl{ConnectionAndFrameCompatibility}$$
which in turn implies that $\Ga=A-\ga K$ is the connection $\Ga(\ep)$ induced by the triad;
see Lemma $(C.6)$ below.

We also remark that using this last result the evolution equation \ShowLabel{SkewFirstOrthogonal}
simplifies to
$$
2\ep^{ADC}\di_{[A} u_{0]} 
= -\ep^{ADC} \tilde K^j \ep_{jA} 
\fn$$

The part orthogonal to $S$ is
$$
\ep^{ABC}\(
2\ga\ep^k_{A} \di_{[B} u_{0]}  
-\ga \si^2\na_A e^k_B
+ 2\ep^k{}_{ij} \na_{[B} e^i_{0]} \ep^j_A
+(\si^2-\ga^2) \si^2\ep^k{}_{ij} K^i_A \ep^j_B\)=0
\fl{sacfdc}$$
By subtraction equations \ShowLabel{sacfdc} and \ShowLabel{SkewFirstOrthogonal} give finally
$$
\ep^{ABC}\(
2\ga\ep^k_{A} \di_{[B} u_{0]}  
+\ep^k{}_{ij} K^i_A \ep^j_B
-\ga \tilde K^j \ep_{jA} \ep^k_B
\)=0
\fn$$
which are $9$ evolutionary equations.

\NewSubSection{Equation for $\de e_0^\mu$}

Let us project the equation 
$$
F^k\land e_k
+(\si^2-\ga^2) \ep_{ijk} K^i\land K^j\land e^k
+2\frac[\si^2-\ga^2/\ga]\na K^i\land e_i
=0
\fn$$
The part parallel to $S$ is
$$
\eqalign{
&\ep^{ABC}\(
F^k_{AB} e_{kC}
+(\si^2-\ga^2) \ep_{ijk} K^i_A K^j_B \ep^k_C
+2\frac[\si^2-\ga^2/\ga]\na_A K^i_B \ep_{iC}
\)=0
\quad \then\cr
&
\ep_k{}^{ij}F^k_{AB} E_i^A E_j^B
+2(\si^2-\ga^2)K^{[i}_A K^{j]}_B E_i^A E_j^B
+2\frac[\si^2-\ga^2/\ga] \na_A K^k_B E_{kC} \ep^{ABC}=0
}
\fn$$
Then by Lemma $(C.9)$ we obtain
$$
\ep_k{}^{ij}F^k_{AB} E_i^A E_j^B
-2(\si^2-\ga^2)K^{[i}_A K^{j]}_B E_i^A E_j^B=0
\fn$$

The part orthogonal to $S$ is
$$
\ep^{ABC}\(
2F^k_{A} e_{kB}
+2(\si^2-\ga^2) \ep_{ijk} \tilde K^i K^j_A \ep^k_B
+4\frac[\si^2-\ga^2/\ga]\na_{[B} K^i_{0]} \ep_{iA}
\)=0
\fn$$
which are evolutionary equations.

\NewSubSection{Equation for $\de e_i^\mu$}

Let us project the equation 
$$
\eqalign{
&2(\si^2-\ga^2) K^i\land e_i\land K^k
+\ep^k{}_{ij} F^i\land e^j
+(\si^2-\ga^2)\na K^k\land e^0 +\cr
&\qquad+\ga(\si^2-\ga^2) \ep^k{}_{ij} K^i\land K^j\land e^0
+\ga F^k\land e^0 
=0
}
\fn$$
The part parallel to $S$ is
$$
\ep^{ABC}\(2(\si^2-\ga^2) K^i_A e_{iB}K^k_C
+\ep^k{}_{ij} F^i_{AB} e^j_C\)
=0
\fn$$
and using \ShowLabel{KeShew} we easily obtain
$$
F^i_{AB} E_i^A
=0
\fl{FourthParallel}$$

The part orthogonal to $S$ is
$$
\eqalign{
\si^2(\si^2-\ga^2) \na_{[A} K^k_{B]}&
-2(\si^2-\ga^2)\tilde K^i K^k_{[A} \ep_{iB]}
+2\ep^k{}_{ij} F^i_{[A} \ep^j_{B]}
+\cr
&+\ga\si^2 (\si^2-\ga^2) \ep^k{}_{ij} K^i_{[A} K^j_{B]}
+ \ga \si^2 F^k_{AB} =0
}
\fn$$
which are $9$ evolutionary equations ($F^i_A$ contains the evolutionary part for the field $A^i_A$).
 
Hence, by finally collecting the constraint equations altogether one gets
$$
\cases{
&\na_A E^A_i=0\cr
&F^i_{AB} E_i^A=0\cr
&\ep_k{}^{ij}F^k_{AB} E_i^A E_j^B
-2(\si^2-\ga^2)K^{[i}_A K^{j]}_B E_i^A E_j^B=0\cr
}
\fn$$
where $\ga K^i_A(A, j^1\ep)=A^i_A-\Ga^i_A$.
These are the starting point of LQG; see \ref{RovelliBook}.

\NewSection{Conclusions and Perpectives}

We produced a spacetime, manifestly covariant Lagrangian formulation of
Ashtekar-Barbero-Immirzi gravity. 
The Lagrangian is simply the pull-back of the Holst's Lagrangian along the 
field variables $(e, A, K)$.

By projecting on a spacelike hypersurface  $S$ we re-obtained in a pretty simple way
the Hamiltonian constraints which are the starting point for the LQG framework together with the
expression of the extrinsic field $K$ in terms of the fields $(E, A)$.
The Lagrangian written in the form \ShowLabel{HBILagrangian} is, to the best of our knowledge, new
in the literature.

We believe that this formulation might provide a better understanding of the gauge covariant structure that LQG is based on. This structure is particularly important since it is the key fact which most of the results of LQG (namely, discretization of space areas and volumes) are based on.
It may also help a better understanding of spin foam models as it provides a spacetime formulation better adapted to the final LQG scheme when compared to the standard Holst's formulation.

Further investigations will be devoted to the general behaviour of a gauge theory endowed with a reduction of the gauge group. 
It is particularly interesting to investigate the behaviour of conservation laws in such a case.

\NewAppendix{\FieldEqsEquivalenceAPP}{Equivalence of Field Equations}

We shall here prove the equivalence between field equations \ShowLabel{BIHFieldsEquations} 
ensuing from the
BIH Lagrangian \ShowLabel{HBILagrangian} and field equations of the usual Holst Lagrangian
\ShowLabel{HolstLagrangian}, namely:
$$
\cases{
&\nab{\om} e^{[a} \land e^{b]} +\frac[\ga/2]\ep_{cd}\^a\^b \> \nab{\om} e^{c} \land e^{d} =0\cr
&R^{ab}\land e_a +\frac[\ga/2]\ep_{cd}\^a\^b \> R^{cd}\land e_a =0\cr
}
\fl{HolstFieldEquations}$$

Let us first notice that
$$
\nab{\om} e^0= \di e^0 + \om^0\_i \land e^i=  \di e^0 +K_i \land e^i
\fn$$
and
$$
\nab{\om} e^i= \di e^i + \om^i\_0 \land e^0 + \om^i\_j \land e^j=
\na e^i +\ga \ep^i{}_{jk} K^j\land e^k +\si^2 e^0\land K^i
\fn$$
where the inverse of transformation \ShowLabel{BIVariables} has been used, namely
$$
\cases{
&\om^{0i}_\mu=K^i_\mu\cr
&\om^{ij}_\mu=\ep^{ij}{}_k \( A^i_\mu -\ga K^i_\mu\)\cr
}
\fn$$

By setting $[ab]=[0i]$ in the first equation of \ShowLabel{HolstFieldEquations} we obtain
$$
\eqalign{
0=& (\di e^0 + K_j\land e^j) \land e^i
-(\na e^i +\ga \ep^i{}_{jk} K^j\land e^k +\si^2 e^0\land K^i)\land e^0+\cr
&+\si^2\ga\ep^i{}_{jk}(\na e^j +\ga \ep^j{}_{lm} K^l\land e^m +\si^2 e^0\land K^j)\land e^k=\cr
=&\di e^0 \land e^i 
-\na e^i\land e^0 
+ \si^2\ga \ep^i{}_{jk} \na e^j\land e^k
+(\si^2-\ga^2)\si^2 K_k\land e^k\land e^i
\equiv \E^{0i}
}
\fn$$

By setting $[ab]=[ij]$ in the first equation of \ShowLabel{HolstFieldEquations} we obtain
$$
\eqalign{
0=& \nab{\om} e^{i} \land e^{j}
-\nab{\om} e^{j} \land e^{i} 
+\ga\ep_{k}{}^{ij} \> \nab{\om} e^{0} \land e^{k}
-\ga\ep_{k}{}^{ij} \> e^{0} \land \nab{\om}  e^{k}
 =\cr
=&
2\na e^{[i}\land e^{j]}
+\ga \ep^k{}^{ij} \di e^0\land e^k
-\ga \ep^k{}^{ij} \na e^k\land e^0
+2(\si^2-\ga^2)e^0\land K^{[i}\land e^{j]}+\cr
&+2\ga\ep_{lm}{}^{[i} e^{j]}\land  e^l\land K^m
+\ga \ep_{k}{}^{ij}  e^k\land K_l\land e^l
\equiv \E^{ij}
}
\fn$$

One can easily check that $\ep^i{}_{jk} \E^{jk}=0$ coincides with the second field equation of
\ShowLabel{BIHFieldsEquations};
analogously the combination $\E^{0i} -\si^2\ga\ep^i{}_{jk} \E^{jk}=0$ coincides with the first
 field equation of  \ShowLabel{BIHFieldsEquations}.
 
By setting $b=0$ in the second equation of \ShowLabel{HolstFieldEquations} we obtain
$$
\eqalign{
0=&R^{0i} \land e^i -\si^2\frac[\ga/2] \ep^i{}_{jk} R^{jk}\land e_i=\cr
=&\si^2\( \ga F^i\land e_i 
-2(\si^2-\ga^2)\na K^i\land e_i
-\ga(\si^2-\ga^2) \ep^i{}_{jk} K^j\land K^k\land e_i
\)
}
\fn$$
where we used \ShowLabel{CurvatureIdentities};
the equation so obtained is equivalent to the third field equation of \ShowLabel{BIHFieldsEquations}.

By setting $b=i$ in the second equation of \ShowLabel{HolstFieldEquations} we obtain
$$
\eqalign{
0=&R^{0i} \land e_0 
-R^{ij} \land e_j
-\si^2\frac[\ga/2] \ep^i{}_{jk} R^{jk}\land e_0
-\ga \ep^{ij}{}_{k} R^{0k}\land e_j
=\cr
=& 2 (\si^2-\ga^2) \na K^i\land e^0
+2(\si^2-\ga^2) k^i\land K^j\land e_j
+\cr&
-\ga(\si^2-\ga^2) \ep^i{}_{jk} K^j\land K^K \land e^0
+\ga F^i\land e^0
+\ep^i{}_{jk} F^j\land e^k
}
\fn$$
which is equivalent to the fourth field equation of \ShowLabel{BIHFieldsEquations}.

\NewAppendix{\FrameAPP}{Projection of the Frame}

In \ref{OurAshtekar} it was shown how one can project (with no gauge fixing) the spacetime tetrads to the
triads on the hypersurface $S\subset M$.

The construction was there adapted to the selfdual formulation based on the projection
on the selfdual bundle
$$
\begindc{\commdiag}[1]
\obj(0,0)[M1]{$M$}
\obj(40,0)[M2]{$M$}
\mor{M1}{M2}{}[\atleft, \solidline] \mor(0,3)(40,3){}[\atleft, \solidline]
\obj(0,40)[P1]{$P$}
\obj(40,40)[P2]{$^+\<\<P$}
\mor{P1}{P2}{}
\mor{P1}{M1}{}
\mor{P2}{M2}{}
\enddc
\fn$$
while we now need to deal with the Barbero-Immirzi formulation that is based on the reduction
$$
\begindc{\commdiag}[1]
\obj(0,0)[M1]{$M$}
\obj(40,0)[M2]{$M$}
\mor{M1}{M2}{}[\atleft, \solidline] \mor(0,3)(40,3){}[\atleft, \solidline]
\obj(0,40)[P1]{$^+\<\<P$}
\obj(40,40)[P2]{$P$}
\mor{P1}{P2}{}
\mor{P1}{M1}{}
\mor{P2}{M2}{}
\enddc
\fn$$

In the discussion about the selfdual formulation we basically resorted to the splitting 
$\Spin(4)\simeq \SU(2)\times \SU(2)$ and to the fact that nothing depends on the antiselfdual part.

In the Barbero-Immirzi case we have a different principal bundle diagram, namely the reduction
$$
\begindc{\commdiag}[3]
\obj(30,10)[S1]{$S$}
\obj(70,10)[S2]{$S$}
\obj(110,10)[S3]{$S$}
\mor{S1}{S2}{}[\atright, \solidline] \mor(30,11)(70,11){}[\atright, \solidline] 
\mor{S2}{S3}{}[\atright, \solidline] \mor(70,11)(110,11){}[\atright, \solidline] 
\obj(50,30)[M1]{$M$}
\obj(90,30)[M2]{$M$}
\obj(130,30)[M3]{$M$}
\mor{M1}{M2}{}[\atright, \solidline] \mor(50,31)(90,31){}[\atright, \solidline] 
\mor{M2}{M3}{}[\atright, \solidline] \mor(90,31)(130,31){}[\atright, \solidline] 
\mor{S1}{M1}{$i$}[\atright, \injectionarrow] 
\mor{S2}{M2}{$i$}[\atright, \injectionarrow] 
\mor{S3}{M3}{$i$}[\atright, \injectionarrow] 
\obj(30,40)[Si1]{$^+\Si$}
\obj(70,40)[Si2]{$\Si$}
\obj(110,40)[Si3]{$L(S)$}
\mor{Si1}{S1}{}[\atright, \solidarrow]
\mor{Si2}{S2}{}[\atright, \solidarrow]
\mor{Si3}{S3}{}[\atright, \solidarrow]
\obj(50,60)[P1]{$^+\<P$}
\obj(90,60)[P2]{$P$}
\obj(130,60)[P3]{$L(M)$}
\mor{P1}{M1}{}[\atright, \solidarrow]
\mor{P2}{M2}{}[\atright, \solidarrow]
\mor{P3}{M3}{}[\atright, \solidarrow]
\mor{Si1}{P1}{$i^\ast$}[\atright, \injectionarrow] 
\mor{Si2}{P2}{$i^\ast$}[\atright, \injectionarrow] 
\mor{P1}{P2}{}[\atright, \solidarrow]
\mor{Si1}{Si2}{}[\atright, \solidarrow]
\mor{P2}{P3}{$\La$}[\atleft, \solidarrow]
\cmor((30,45)(40,50)(70,52)(100,50)(110,45)) \pdown(60,55){$\tilde\La$}[\atleft, \solidarrow] 
\enddc
\fn$$
Moreover, in the Lorentzian case it is non-trivial to find a group of spin transformations fixing the BI connection.
We hence revert to another argument, which remarkably leads to the same result.

Let us start with a point $(k, {}^+\<S)\in {}^+\<\Si$;
it induces a point $(i(k), {}^+\<S)\in {}^+\<P$ and a point $(i(k), {}^+\<S, {}^+\<S)\in P$.
Still there exists a unique element $(\one, {}^-\<S)\in\Spin(4)$ such that
$\ell(\one, {}^-\<S)(e_0)=u$; hence the frame $\hat e_a=e_a \cdot \ell((\one, {}^-\<S)$ is adapted to the submanifold $S\subset M$.
If the fields $(e_a^\mu, \te^{ab}_\mu)$ provide a solution for the tetrad-affine formalism,
then the corresponding $(\hat e_a^\mu, A^i_\mu, K^i_\mu)$ provide a solution of the Barbero-Immirzi formulation.

The Lorentzian case goes as the Euclidean one, except that the reduction is obtained by the group morphism
$i:\Spin(3)\simeq \SU(2)\arr \Spin(3,1)\simeq \SL(2, \C): u\mapsto u$.
Hence one proves that there exists a unique element in the form
$S=\al+E v\in \Spin(3,1)$ with $\al\in \R^+$ such that the frame $\hat e_a=e_a \cdot \ell((\one, {}^-\<S)$ is adapted to the submanifold $S\subset M$.
If the fields $(e_a^\mu, \te^{ab}_\mu)$ provide a solution for the tetrad-affine formalism,
then the corresponding $(\hat e_a^\mu, A^i_\mu, K^i_\mu)$ provide a solution of the Barbero-Immirzi formulation.

In both cases the triad so obtained transforms as expected for a spin frame on $^+\<\Si$.

\NewAppendix{\ProjectionIdentitiesAPP}{Projection onto a Hypersurface}

We shall collect here few tricks used in the above derivation of projected equations.

Let us consider a (possibly local) $1$-form $\al=\al_\mu\>\di x^\mu$; 
we can define two fields on $S$
$$
\cases{
&\al_A(k) := \al_\mu(x(k)) \del_A x^\mu(k)\cr
&\tilde \al(k) := \al_\mu(x(k)) u^\mu(k )\cr
}
\fl{1formProjection}$$
These two fields are a $1$-form and a scalar over $S$, respectively.

One can do the same with a $2$-form $\be=\be_{\mu\nu}\>\di x^\mu\land \di x^\nu$; 
we can define two fields on $S$
$$
\cases{
&\be_{AB}(k) := \be_{\mu\nu}(x(k)) \del_A x^\mu(k)\del_B x^\nu(k)\cr
&\tilde \be_A(k) := \be_{\mu\nu}(x(k)) u^\mu(k) \del_A x^\nu(k)\cr
}
\fn$$
These two fields are a $2$-form and a $1$-form over $S$, respectively.

By projecting the differential $\be=\di\al$, one obtains
$$
\cases{
&\be_{AB} := \di_{[\mu} \al_{\nu]} \del_A x^\mu\del_B x^\nu\cr
&\tilde \be_A := \di_{[\mu} \al_{\nu]} u^\mu \del_A x^\nu\cr
}
\fn$$

By differentiating the first equation of \ShowLabel{1formProjection}  with respect to $k^B$,
we easily get
$$
\di_{[B}\al_{A]} = \di_{[\nu} \al_{\mu]}\del_A x^\mu\del_B x^\nu=\be_{AB}
\fl{ParallelDifferentialID}$$ 
which means that the part of the differential $\di \al$ which is parallel to $S$ is in fact the differential (on $S$) of 
the parallel part to $S$ of $\al$ itself.

We stress that the same trick does not hold for the orthogonal part; the orthogonal part of the differential
cannot be computed on $S$ alone. This means that, as expected, the differential in the orthogonal direction does somehow encode {\it evolution} of fields on $S$ (or better it {\it would} encode evolution
if we considered a foliation). 
Nevertheless, we shall call {\it evolutionary} the equations involving the orthogonal part of the differential.

Let us thus consider the special case in which $\al_\mu\equiv e^0_\mu$.
In this case we obtain $\al_A=0$ and $\tilde \al=\si^2$; 
by applying \ShowLabel{ParallelDifferentialID} we obtain directly
$$
\di_{[A} u_{B]}\equiv\di_{[A} e^0_{B]}=0
\fn$$

Let us now prove the following lemmas.

\LEMMAl{TorsionLemma}
There exists a unique connection $\Ga^i_A$, namely the connection $\Ga^i_A(\ep)$ induced by the triad,
such that $\nab{\Ga}_{[A} \ep^{i}_{B]}=0$.
\ENDLEMMA

\PROOF
Let us introduce the quantity $\De^i_m= (\Ga^i_A-\Ga^i_A(\ep) )\ep^A_m=: \frac[1/2] \ep^i{}_{jk} \De^{jk}{}_m$.
The hypothesis can be written in term of this quantity as
$$
\De^i{}_{[km]}=0
\fn$$
Hence we obtain a quantity $\De_{ijk}$ which is antisymmetric in $[ij]$ and symmetric in $(jk)$.
Then one trivially proves
$$
\De_{ijk}= -\De_{jik}=- \De_{jki}= \De_{kji}= \De_{kij}= -\De_{ikj}=-\De_{ijk}
\quad\then\quad \De_{ijk}=0
\fn$$
which in turn proves that $\Ga^i_A=\Ga^i_A(\ep)$.
\ENDPROOF

\LEMMAl{naKLemma}
$\na_A K^k_B E_{kC} \ep^{ABC}=-2\ga K^{[i}_A K^{j]}_B E_i^A E_j^B$
\ENDLEMMA

\PROOF
We have:
$$
\eqalign{
\na_A K^i_B \ep_{iC} \ep^{ABC}=&
\di_A K^i_B \ep_{iC} \ep^{ABC} -\ep^i{}_{jk} A^j_A K^k_B \ep_{iC} \ep^{ABC}=\cr
=&- K^i_B \di_A \ep_{iC} \ep^{ABC} -\ep^i{}_{jk} A^j_A K^k_B \ep_{iC} \ep^{ABC}\Label={ConnectionAndFrameCompatibility}\cr
=&- K^i_B \ep_{ilm}\Ga^l_A\ep^m_C \ep^{ABC} -\ep^i{}_{jk} A^j_A K^k_B \ep_{iC} \ep^{ABC}=\cr
=&- K^i_B \ep_{ilm}(\ga K^l_A - A^l_A)\ep^m_C \ep^{ABC} -\ep^i{}_{jk} A^j_A K^k_B \ep_{iC} \ep^{ABC}=\cr
=&-2\ga K^{[i}_A K^{j]}_B   E_i^A \ep_j^B\cr
}
\fn$$
from which the Lemma readily follows.
\ENDPROOF

\Acknowledgements

This work is partially supported by MIUR: PRIN 2005 on ``Leggi di conservazione e termodinamica in meccanica dei continui e teorie di campo''.  We also acknowledge the contribution of INFN (Iniziativa Specifica NA12) and the local research founds of Dipartimento di Matematica of Torino University.

\

\ShowBiblio


\end